\documentclass[12pt]{article}

\usepackage{sbc-template}
\usepackage{graphicx,url}
\usepackage[utf8]{inputenc}
\usepackage[T1]{fontenc}
\usepackage[english]{babel}
\usepackage{subcaption}
\usepackage{fontawesome}
\usepackage{amsmath} 
\usepackage{float}
\usepackage[nolist]{acronym}
\usepackage{fancybox}
\usepackage{enumitem}
\usepackage{array} 
\usepackage{booktabs} 
\usepackage[table]{xcolor}

\usepackage{tikz}

\title{ARCADE: A RAN Diagnosis Methodology in a Hybrid AI Environment for 6G Networks}

\author{Daniel Ricardo {Cunha Oliveira}\inst{1}, Rodrigo Moreira\inst{1,2}\\Flávio {de Oliveira Silva}\inst{1,3}}

\address{
  Faculty of Computing (FACOM) -- Federal University of Uberlândia (UFU)\\
  Uberlândia -- MG -- Brazil
\nextinstitute
  Federal University of Viçosa
  (UFV)\\
  Rio Paranaíba -- MG -- Brazil
\nextinstitute
  Department of Informatics -- School of Engineering\\
  University of Minho -- Braga, Portugal
  \email{drcoliveira@ufu.br, rodrigo@ufv.br, flavio@di.uminho.pt}
}

\begin{document} 

\begin{acronym}

\acro{3GPP}{Third Generation Partnership Project}
\acro{5GC}{5G Core}

\acro{ALU}{Arithmetic Logic Unit}
\acro{ANPM}{All Name Prefix Matching}
\acro{API}{Application Programming Interface}
\acro{ASIC}{Application-Specific Integrated Circuit}
\acro{ASNM}{All Sub-Name Matching}
\acro{AI}{Artificial Intelligence}
\acro{ANN}{Artificial Neural Network}
\acro{AR}{Augmented Reality}
\acro{ARCADE}{Automated Radio Coverage Anomalies Detection and Evaluation}
\acro{AIC}{Advanced Interference Cancellation}
\acro{AMF}{Access and Mobility Management Function}
\acro{AI-Agent}{Artificial Intelligence Agent}

\acro{BGP}{Border Gateway Protocol}
\acro{BM}{Bypass Memory}
\acro{BMv2}{Behavioral Model Version 2}
\acro{BS}{Base Station}
\acro{BTS}{Base Transceiver Station}
\acro{BBU}{Baseband Unit}

\acro{CAD}{Component Action Data}
\acro{CAM}{Content Addressable Memory}
\acro{CFIB}{Control Plane Forwarding Information Base}
\acro{CNAME}{Canonical Name}
\acro{CoFIB}{Compressed Forwarding Information Base}
\acro{CPE}{Canonical Prefix Extractor}
\acro{CPU}{Central Processing Unit}
\acro{CS}{Content Store}
\acro{CSw}{Core Switch}
\acro{CPST}{Control Plane Shape Table}
\acro{CNN}{Convolutional Neural Network}
\acro{CNNs}{Convolutional Neural Networks}
\acro{CQI}{Channel Quality Indicator}
\acro{CN}{Core Network}
\acro{CUPS}{Control- and User-Plane Separation}
\acro{CoMP}{Coordinated Multi-Point}
\acro{C/I}{Carrier-to-Interference Ratio}
\acro{CA}{Coverage Area}
\acro{CI}{Coverage Index}
\acro{CQualI}{Cell Quality Index}

\acro{DCH}{Dual Component Hashmap}
\acro{DFIB}{Dataplane FIB}
\acro{DNS}{Domain Name System}
\acro{DPST}{Data Plane Shape Table}
\acro{DRAM}{Dynamic Random Acess Memory}
\acro{dRMT}{disaggregated Reconfigurable Match-Action Table}
\acro{DC}{Dual Connectivity}
\acro{DLT}{Distributed Ledger Technology}
\acro{DRL}{Deep Reinforcement Learning}
\acro{DT}{drive test}
\acro{dBm}{decibéis-miliwatt}
\acro{dB}{decibel}
\acroplural{dB}[dBs]{decibéis}

\acro{ESw}{Edge Switch}
\acro{ENM}{Exact Name Matching}
\acro{E2E}{End-to-End}
\acro{ERB}{Estação Rádio-Base}
\acroplural{ERB}[ERBs]{Estações Rádio-Base}
\acro{eNWDAF}{Enhanced Network Data Analytics Function}
\acro{EMS}{Element Management System}
\acro{E-UTRAN}{Evolved Universal Terrestrial Radio Access Network}
\acro{EPC}{Evolved Packet Core}
\acro{EIRP}{Effective Isotropic Radiated Power}
\acro{eICIC}{Enhanced Inter-Cell Interference Coordination}

\acro{FANTNet}{Fast NDN Transit Networking}
\acro{FIB}{Forwarding Information Base}
\acro{FIFO}{First-In First Out}
\acro{FPGA}{Field-Programmable Gate Array}
\acro{FST}{Fast Switching Table}
\acro{FR2}{Frequency Range 2}
\acro{FeICIC}{Further Enhanced Inter-Cell Interference Coordination}

\acro{GPU}{Graphics Processing Unit}
\acro{GP}{Gaussian Process}
\acro{GPSK}{Gaussian Process with Spatial Kernel}
\acro{GNSS}{Global Navigation Satellite System}
\acro{GERAN}{GSM EDGE Radio Access Network}

\acro{HBM}{Hashtray-Based Method}
\acro{HCT}{Hash Conflicting Table}
\acro{HTTP/2}{Hypertext Transfer Protocol version 2}

\acro{ICN}{Information-Centric Networking}
\acro{IP}{Internet Protocol}
\acro{IoT}{Internet of Things}
\acro{IA}{Inteligência Artificial}
\acro{ICIC}{Inter-Cell Interference Coordination}
\acro{ICI}{Inter-Cell Interference}
\acro{IBN}{Intent-based Networking}
\acro{IAX}{Interference Affected Index}
\acro{ISI}{Interference Source Index}


\acro{KPI}{Key Performance Indicator}

\acro{LFU}{Least Frequently Used}
\acro{LIB}{Label Information Base}
\acro{LNPM}{Longest Name Prefix Matching}
\acro{LPM}{Longest Prefix Matching}
\acro{LRU}{Least Recently Used}
\acro{NPC}{NDN Packet Converter}
\acro{LTE}{Long Term Evolution}
\acro{LPWA}{Low-Power Wide Area}

\acro{MPLS}{Multi Protocol Label Switching}
\acro{MLaaS}{Machine Learning as a Service}
\acro{MAG}{Monitoring Agent}
\acro{MAGs}{Monitoring Agents}
\acro{MDT}{Minimization of Drive Tests}
\acro{ML}{Machine Leaning}
\acro{MC}{Multi-Connectivity}
\acro{MLO}{Multi-link Operation}
\acro{MR}{Measurement Report}
\acro{MSE}{Mean Squared Error}
\acro{MANO}{Management and Orchestration}
\acro{MAC}{Medium Access Control}
\acro{MSE}{Mean Squared Error}

\acro{NCH}{Name Component Hash Table}
\acro{NDN}{Named Data Networking}
\acro{NIC}{Network Interface Card}
\acro{NLSR}{Named-data Link State Routing}
\acro{NFV}{Network Function Virtualization}
\acro{NIC}{Network Interface Card}
\acro{NN}{Neural Network}
\acro{MEC}{Multi-Access Edge Computing}
\acro{NS}{Network Slices}
\acro{NWDAF}{Network Data Analytics Function}
\acro{NMS}{Network Management System}
\acro{MCS}{Mission-Critical Service}
\acro{MPS}{Mission Priority Service}
\acro{NG-RAN}{Next-Generation Radio Access Network}
\acro{NSSF}{Network Slice Selection Function}
\acro{NAS}{Non-Access Stratum}
\acro{NF}{Network Function}
\acro{NGAP}{NG Application Protocol}
\acro{NE}{Network Element}

\acro{OVS}{Open vSwitch}
\acro{OI}{Overlap Index}

\acro{P4}{Programming Protocol-Independent Packet Processors}
\acro{P4NF}{P4 Name Friendly Format}
\acro{PDU}{Packet Data Unit}
\acro{PISA}{Protocol-Idependent Switch Architecture}
\acro{PIT}{Pending Interest Table}
\acro{PNA}{Portable NIC Architecture}
\acro{PSA}{Portable Switch Architecture}
\acro{PCI}{Physical Cell ID}
\acro{PD}{Packet Duplication}
\acro{PRB}{Physical Resource Block}
\acro{PCF}{Policy Control Function}

\acro{QoS}{Quality-of-Service}
\acro{QUIC}{Quick UDP Internet Connection}
\acro{RIB}{Routing Information Base}
\acro{RMT}{Reconfigurable Match Tables}
\acro{RTT}{Round Trip Time}
\acro{RAN}{Radio Access Network}
\acro{RANs}{Radio Access Networks}
\acro{RBF}{Radial Basis Function}
\acro{RF}{radiofrequency}
\acro{RL}{Reinforcement Learning}
\acro{RMSE}{Root Mean Squared Error}
\acro{RNN}{Recurrent Neural Network}
\acro{RSRP}{Reference Signal Received Power}
\acro{RSRQ}{Reference Signal Received Quality}
\acro{RLC}{Radio Link Control}
\acro{RRC}{Radio Resource Control}
\acro{RSSI}{Received Signal Strength Indicator}
\acro{RNA}{Rede Neural Artificial}
\acro{RBS}{Radio Base Station}

\acro{PDCP}{Packet Data Convergence Protocol}

\acro{SDN}{Software Defined Networking}
\acro{SRAM}{Static Random Access Memory}
\acro{SFI2}{Slicing Future Internet Infrastructures}
\acro{SL}{Single-Link}
\acro{SON}{Self-Organizing Networks}
\acro{SBA}{Service-Based Architecture}
\acro{SBI}{Service-Based Interface}
\acro{SINR}{Signal-to-Interference-plus-Noise Ratio}

\acro{TCAM}{Ternary Content Addressable Memory}
\acro{TCP}{Transmission Control Protocol}
\acro{TLS}{Transport Layer Security}
\acro{TLV}{Type-Length Value}
\acro{TNA}{Tofino Native Architecture}
\acro{TN}{Transport Network}

\acro{URI}{Uniform Resource Identifier}
\acro{URL}{Uniform Resource Locator}
\acro{UE}{User Equipment}
\acro{URLLC}{Ultra-Reliable Low-Latency Communication}
\acro{UTRAN}{Universal Terrestrial Radio Access Network}
\acro{UMTS}{Universal Mobile Telecommunication System}

\acro{VLSI}{Very Large Scale Integrated}
\acro{VNF}{Virtual Network Function}



\end{acronym}
\maketitle

\begin{abstract}

\ac{AI} plays a key role in developing 6G networks. While current specifications already include \ac{NWDAF} as a network element responsible for providing information about the core, a more comprehensive approach will be needed to enable automation of network segments that are not yet fully explored in the context of 5G. In this paper, we present \ac{ARCADE}, a methodology for identifying and diagnosing anomalies in the cellular access network. Furthermore, we demonstrate how a hybrid architecture of network analytics functions in the evolution toward 6G can enhance the application of \ac{AI} in a broader network context, using \ac{ARCADE} as a practical example of this approach.

\end{abstract}
     
\section{Introduction}\label{sec:introduction}

In recent years, cellular networks have become an indispensable and essential component of society, particularly in a globalized digital community that relies on ubiquitous and reliable services. The exponential increase in connected devices has made \acp{RAN} highly complex, especially considering the coexistence of multiple technology generations and the growing demand for \ac{IoT}-based solutions. In this context, there is an increasing need for solutions that simplify \ac{RAN} optimization, particularly since, in systems such as 4G and 5G—and potentially in 6G, should the next generation adopt an evolutionary approach to the access network—network self-interference is inversely correlated with efficiency and traffic capacity. Traditional network optimization methods, which rely on manual processes and highly specialized technical expertise, are becoming increasingly ineffective as system capacity and complexity expand.  

This study presents \ac{ARCADE}, a methodological approach for evaluating and identifying coverage anomalies in a standard 3GPP cellular system. More importantly, we address a key issue: how \ac{ARCADE} acquires data from the \ac{RAN} to analyze coverage and interference among cells within a specific cluster under investigation. This is achieved through a hybrid architecture proposal for implementing \ac{AI} in the future 6G core network.

This work is organized as follows: Section~\ref{sec:related_work} presents an overview of related work in this area. Section~\ref{sec:arcade} describes \ac{ARCADE}. Section~\ref{sec:enwdaf} contextualizes an \ac{NWDAF} evolution. Section~\ref{sec:arcadeenwdaf} correlates \ac{ARCADE} and the \ac{NWDAF} evolution. Section~\ref{sec:conclusoes} concludes this study.

\section{Related Work}
\label{sec:related_work}

To provide analytics services in mobile networks, 3GPP has defined \ac{NWDAF} \cite{3GPP_TS_23501} as the central \ac{AI} element in the 5G core topology. 

\begin{figure}  
   \centering  
   \includegraphics[width=0.8\textwidth]{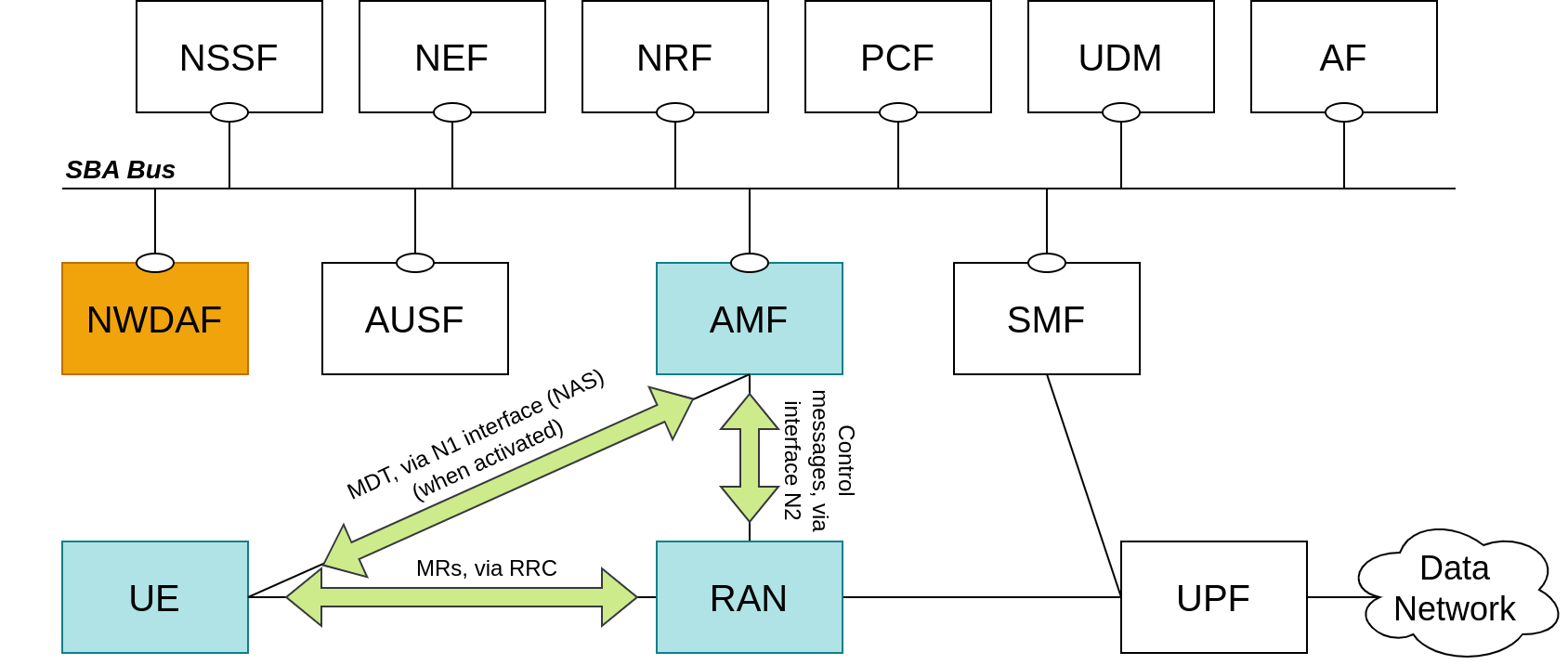}  
   \caption{Simplified signaling diagram of \acp{MR} and \ac{MDT} between network elements. Figure created by the authors.}  
   \label{fig:sinalizacao_nwdaf}  
\end{figure}

\ac{NWDAF} is integrated into the \ac{5GC} architecture as an independent network function, interacting with other network functions through standardized 3GPP interfaces. Although some researchers claim that \ac{NWDAF} is responsible for \ac{AI} and machine learning tasks, it was not initially designed for such purposes and is instead restricted to analytics services \cite{wu_toward_2021}. The current standardized architecture does not allow \ac{NWDAF} to access raw \ac{RF} data, particularly \acp{MR}, which contains rich information about coverage, neighboring cells, and network quality. This limitation arises from the centralized deployment of this network element, as illustrated in Figure~\ref{fig:sinalizacao_nwdaf}, where \ac{NWDAF} connects to other \acp{NF} solely through the \ac{SBA} message bus.  

Thus, a key challenge lies in developing a solution that enables \ac{RAN} data to be stored and potentially processed by the network architecture's analytics elements, particularly \ac{NWDAF}.

A hybrid approach to evolving an architecture that enables the use of \ac{AI} in next-generation networks, as described in this study, was proposed in \cite{souza_neto_w6g_2024}, contrasting with implementation perspectives based on centralized (\cite{abbas_ensemble_2022}) or distributed architectures (\cite{9730220}). Due to space constraints, a detailed overview of the proposed hybrid approach for the \ac{NWDAF} cannot be included here. However, a comprehensive discussion is available in \cite{souza_neto_w6g_2024}.

Furthermore, several studies have proposed approaches similar to \ac{ARCADE}. Some efforts focus on modeling digital twins for mobile networks and \ac{RF} environments in general, as presented in \cite{deng_digital_2021} and \cite{nguyen_digital_2021}. However, these approaches are not exclusively focused on \ac{RAN} or do not aim to model cellular coverage. None of the reviewed studies exhibits all the ideal characteristics within the proposed context. Therefore, we contrast our work with the existing literature in Table~\ref{tab:related_work}, highlighting essential features and their differences. Small dots indicate that a given criterion does not apply to the respective study.

\begin{table}[]
\centering
\scriptsize
\caption{Short State-of-the-Art Survey.}
\label{tab:related_work}
\resizebox{\textwidth}{!}{%
\begin{tabular}{lcccccc}
\hline
\rowcolor[HTML]{EFEFEF} 
\multicolumn{1}{c}{\cellcolor[HTML]{EFEFEF}\textbf{Work}} & \textbf{\begin{tabular}[c]{@{}c@{}}ML/Neural\\ network-based\end{tabular}} & \textbf{\begin{tabular}[c]{@{}c@{}}Independent of\\ project data\end{tabular}} & \textbf{\begin{tabular}[c]{@{}c@{}}Independent of\\ analytical prediction\end{tabular}} & \textbf{\begin{tabular}[c]{@{}c@{}}Supports\\ coverage anomalies\end{tabular}} & \textbf{\begin{tabular}[c]{@{}c@{}}Independent of\\ system KPIs\end{tabular}} & \textbf{\begin{tabular}[c]{@{}c@{}}Applies to 3GPP\\ systems (RAN)\end{tabular}} \\ \hline

Juan Deng \textit{et} al. 2021~\cite{deng_digital_2021} & \faCircleO & \faCircleO & \faCircleO & \faCircleO & \faCircleO & \faCircle \\

\rowcolor[HTML]{EFEFEF} 
MingXue Wang \textit{et} al. 2018~\cite{wang_anomaly_nodate} & \faCircle & \faCircle & \faCircleO & \faCircle & \faCircle & \faCircle \\

Anita Gehlot \textit{et} al. 2022~\cite{gehlot_application_2022} & \faCircle & \textperiodcentered & \textperiodcentered & \textperiodcentered & \faCircleO & \faCircle \\

\rowcolor[HTML]{EFEFEF} 
Kirkwood \textit{et} al. 2022~\cite{kirkwood_bayesian_2022} & \faCircle & \textperiodcentered & \faCircle & \textperiodcentered & \textperiodcentered & \faCircleO \\

Skocaj \textit{et} al. 2022~\cite{skocaj_cellular_2022} & \faCircle & \faCircleO & \faCircleO & \faCircle & \faCircleO & \faCircle \\

\rowcolor[HTML]{EFEFEF} 
Nguyen, 2021~\cite{nguyen_digital_2021} & \textperiodcentered & \textperiodcentered & \textperiodcentered & \textperiodcentered & \textperiodcentered & \textperiodcentered \\

Zhou, 2023~\cite{zhou_gaussian_2023} & \faCircle & \textperiodcentered & \faCircle & \textperiodcentered & \textperiodcentered & \faCircleO \\

\rowcolor[HTML]{EFEFEF} 
Tang, 2023~\cite{tang_multi-output_2023} & \faCircle & \faCircleO & \faCircle & \faCircle & \faCircle & \faCircleO \\

Carmack, 2021~\cite{carmack_neural_2021} & \faCircle & \textperiodcentered & \textperiodcentered & \textperiodcentered & \textperiodcentered & \faCircle \\

\rowcolor[HTML]{EFEFEF} 
Cheerla, 2018~\cite{cheerla_neural_2018} & \faCircle & \faCircleO & \faCircleO & \textperiodcentered & \faCircle & \faCircle \\

Dreifuerst, 2021~\cite{dreifuerst_optimizing_2021} & \faCircle & \faCircleO & \faCircle & \faCircle & \textperiodcentered & \faCircle \\

\rowcolor[HTML]{EFEFEF} 
Ojo, 2020~\cite{ojo_radial_2021} & \faCircle & \faCircleO & \faCircle & \textperiodcentered & \textperiodcentered & \faCircle \\

\textbf{This work} & \faCircle & \faCircle & \faCircle & \faCircle & \faCircle & \faCircle \\
\hline
\end{tabular}%
}
\end{table}

\section{The ARCADE Methodology}
\label{sec:arcade}

In this section, we briefly describe the \ac{ARCADE} methodology, an approach for detecting and evaluating anomalies in the coverage of a cellular system. Due to the limited space available for publication, we emphasize that this work does not aim to provide a detailed description of \ac{ARCADE} -- an endeavor that will be addressed in future articles. Instead, our objective is to leverage \ac{ARCADE}’s framework to demonstrate how a hybrid architecture integrating \ac{AI} elements within a 6G core could enhance applications that currently lack access to critical data due to the constraints of the existing 5G network topology.

\subsection{Proposal}

\ac{ARCADE} introduces a methodology for the automated identification of cellular coverage anomalies, relying solely on georeferenced cell coverage level data, or \ac{RSRP}, without requiring information from \ac{RBS} design plans, network management system \acp{KPI}, or geographic and morphological databases—resources commonly used in most existing approaches \cite{wang_anomaly_nodate} \cite{skocaj_cellular_2022} \cite{dreifuerst_optimizing_2021} \cite{ojo_radial_2021}. Moreover, the coverage data may be sparse and not fully encompass the analyzed area, necessitating coverage extrapolation based on the available data. In addition to the absence of design data, \ac{ARCADE} does not employ mathematical prediction models, excluding, for instance, classical approaches such as Okumura, Hata, or COST-231 \cite{singh_comparison_2012}. Eliminating these dependencies simplifies the solution, given the complexity of obtaining reliable and up-to-date field data.  

\subsection{Data Acquisition}
\label{sec:arcade/aquisicao}  

\begin{figure}  
   \centering  
   \includegraphics[width=0.6\textwidth]{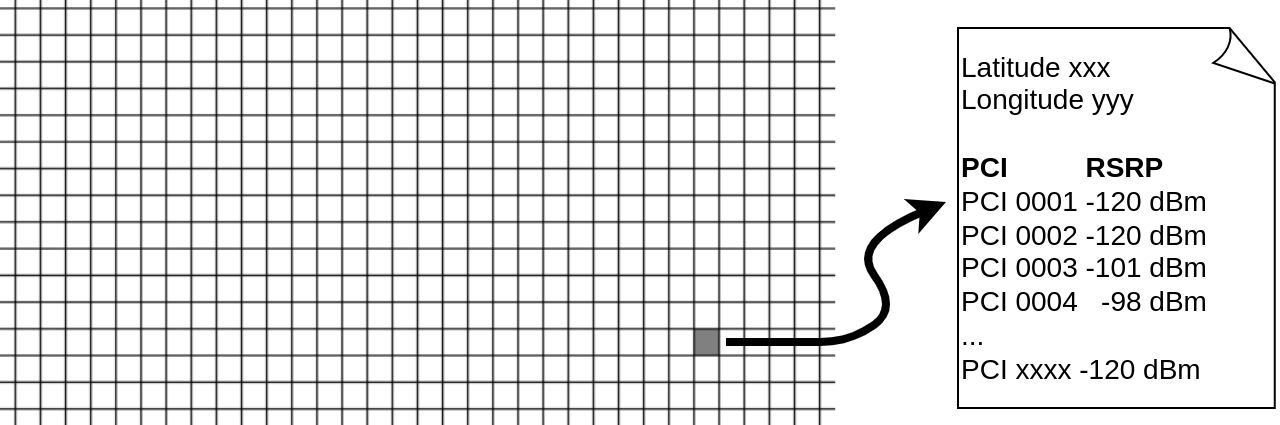}  
   \caption{Georeferenced grid structure and information per grid element. Figure created by the authors.}  
   \label{fig:geogrid}  
\end{figure}

\ac{ARCADE} takes as input \ac{RSRP} samples from various cell \acp{PCI} within the analyzed cluster. Additionally, the samples must be georeferenced, meaning they contain geographic coordinates indicating where each measurement was collected. These coordinates position the samples within a grid that maps the entire area of interest. Each grid element stores information on \ac{PCI} and \ac{RSRP}, as illustrated in Figure~\ref{fig:geogrid}.  

Acquiring these samples is a challenge in itself. Several sources can be used to collect \ac{RF} environment data and obtain information about the state of coverage and interference in a mobile system. One option is \ac{DT}, which involves using probes on vehicles to capture and decode \ac{RBS} signals, storing the data along with geolocation obtained via a \ac{GNSS}. More recently, however, data acquisition has become possible through the 3GPP-defined \ac{MDT} functionality \cite{3GPP_37320}. This feature allows the cellular network to collect \acp{MR}, including georeferenced data, using modern smartphones' built-in \ac{GNSS} capabilities. These devices provide critical \ac{RF} measurement data, such as cell identification via \ac{PCI} and \ac{RSRP}, among other metrics, offering a valuable resource for mapping the radio environment of a cellular system. However, the industry adoption of \ac{MDT} remains limited, and additional constraints on its usage persist, making the optimal solution for acquiring \ac{RF} environment data in a cellular system an open problem.  




\subsection{Coverage Extrapolation}  

Conceptually, \ac{ARCADE} requires a complete \ac{RF} data table for each \ac{PCI} within the grid covering the area of interest. Since not all cells have measurements in every grid element, coverage extrapolation for each cell is necessary. This generalization must model coverage as accurately as possible, considering both anomalous coverage samples (which may indicate suboptimal coverage configurations, such as improper antenna positioning) and outliers caused by device measurement errors (which should be discarded). The potential confusion between coverage anomalies and measurement errors necessitates a sample classification methodology. Once the samples are classified, the total grid area is divided into \emph{normal} and \emph{abnormal} regions, referring to areas where coverage is expected versus areas where coverage results from anomalies such as cell overshooting.  

Once each cell's normal and abnormal regions are defined, sample augmentation is applied to the coverage anomaly areas. This augmentation serves two purposes: first, to establish boundary conditions that prevent anomalies from causing the effect of overfitting in the subsequent modeling with \ac{ANN}; second, to emphasize the impact of anomalies so that the modeling process effectively captures potential overshooting effects. Our approach relies on data extrapolation using Gaussian processes with a spatial kernel, specifically the \ac{RBF} kernel \cite{scholkopf_learning_2002}.  

Once a sufficient number of samples allows for characterizing elements outside the normal coverage area—highlighting anomaly regions and establishing boundary conditions in areas distant from the cell’s primary coverage—it is proposed that an \ac{ANN} be trained with this data to generically model coverage across all grid elements in the total area. The structure of the \ac{ANN} should be determined through experiments evaluating its efficiency and accuracy in generalizing the model. After this sequence of processes, the entire \ac{RF} environment is ultimately described, enabling the classification of cells based on the normality of their coverage.

\subsection{Coverage Anomaly Identification}  

Finally, once the final dataset to be evaluated is obtained, containing the information shown in Figure \ref{fig:geogrid}, each cell's coverage and interference assessment phase begins. This phase determines the cell interaction and how the system could be optimized by increasing or reducing their coverage. It is impossible to analyze a single cell's coverage without considering the context of other cells within the cluster. Even in a highly efficient and optimized system, interference between neighboring cells is inevitable. There is a minimum acceptable level of overlap between adjacent cells, below which coverage adjustments to reduce interference would degrade service quality. Therefore, it is necessary to define measurable parameters for evaluating both the coverage of a cell within the context of a cluster and the interference among them.  

To this end, indicators are defined and associated with the \acp{PCI}, characterizing their coverage-related attributes, such as \emph{Coverage Index}, \emph{Interfering Index}, \emph{Interfered Index}, \emph{Overlap Index}, \emph{Coverage Quality Index}, and \emph{Coverage Matrix}. Analyzing these indicators and their correlations makes it possible to identify anomalous cells and diagnose the \ac{RAN} in a cellular system.  

\section{Evolved NWDAF and the Hybrid Approach}  
\label{sec:enwdaf}  

\begin{figure}  
   \includegraphics[width=\textwidth]{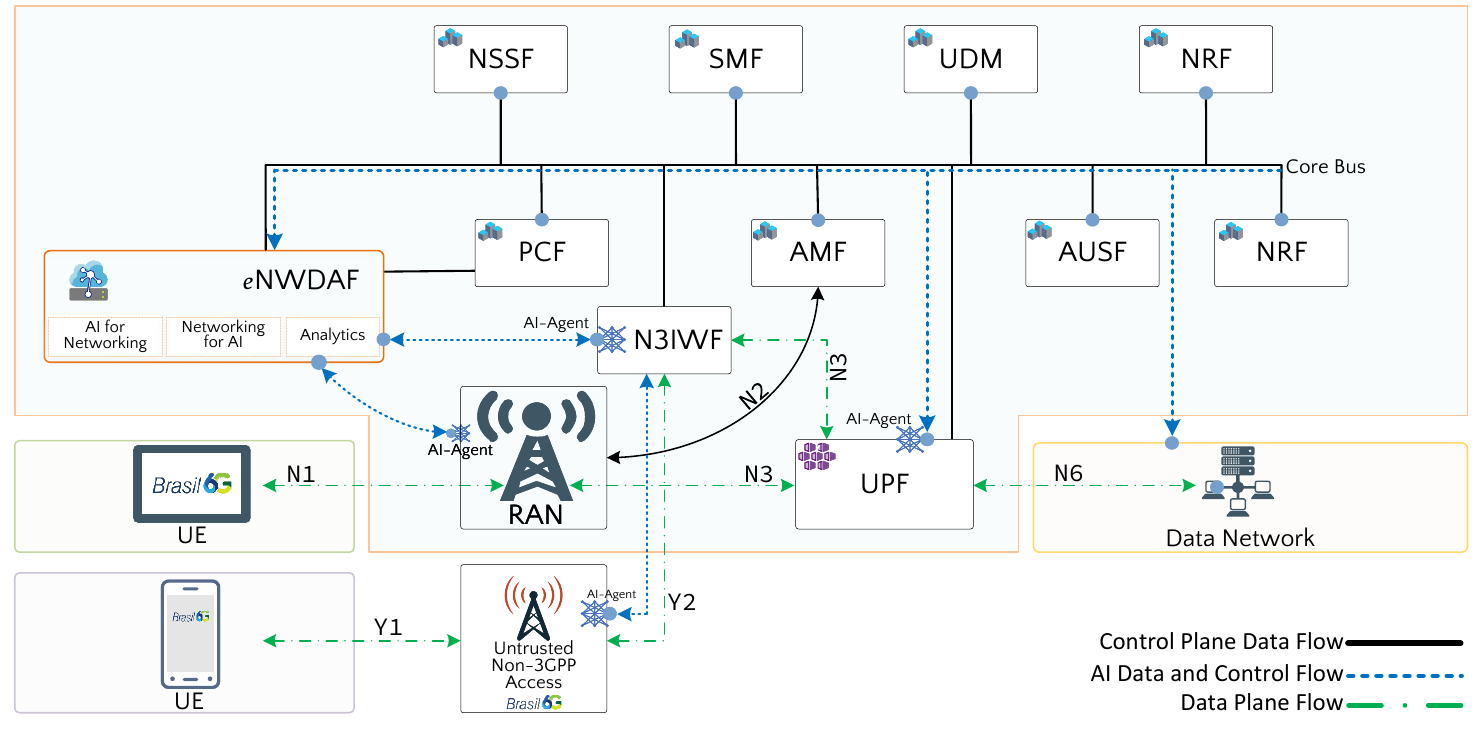}  
   \caption{Advancements in \ac{AI}-Related Core: The \ac{eNWDAF} in 6G Architecture. Source: ~\cite{souza_neto_w6g_2024}}  
   \label{fig:proto6g_core_enhancement}  
\end{figure}

The hybrid architecture for \ac{NWDAF} and the definition of \ac{eNWDAF} were proposed in \cite{souza_neto_w6g_2024}. The architecture introduces \emph{AI-Agents}, capable of collecting data from the \ac{RAN} and other network points that are currently inaccessible to \ac{NWDAF} as specified today. This enhancement expands the role of \ac{NWDAF}, integrating broader \ac{AI} functions in terms of network topology reach. The \ac{eNWDAF} provides \emph{AI for Networking}, \emph{Networking for AI}, and \emph{analytics} support within a single component. Its ability to access network elements beyond the core and its positioning within the \ac{SBA} bus characterize it as a hybrid architecture. Figure \ref{fig:proto6g_core_enhancement} presents a conceptual illustration of this approach.  

\section{ARCADE in the Context of the Hybrid AI Approach}  
\label{sec:arcadeenwdaf}  

An example of \ac{ARCADE} application within the new architecture illustrates the concepts proposed above. In this specific case, two key types of data may be collected: \acp{MR} and \ac{MDT} data.  

As previously mentioned, \acp{MR} data are transmitted from the \ac{UE} to the \ac{RAN} and remain inaccessible to the \ac{NWDAF} in the current 5G architecture. In the proposed architecture, the \acp{RBS} elements, which constitute the \ac{RAN}, incorporate an \emph{AI-Agent} in their baseband processing units. These agents would collect, anonymize, classify, and consolidate the measurement data received. Once the data are organized adequately into ``packets'', they can be transmitted to the \ac{eNWDAF} for centralized analysis by \ac{ARCADE}. This organization may include, for instance, the consolidation of measurements per \ac{UE} over a short period, enabling the characterization of a measurement as a specific sample from a geographical point in the \ac{RF} environment.  

The transmission of \ac{MDT} data is already standardized by 3GPP. However, the volume of available information may be reduced due to limited compatibility of \acp{UE} with the technology among the user base or due to 3GPP specification requirements, which mandate user consent for data collection. Although the available data volume may be lower, \ac{MDT} data can serve as a highly valuable training base for \ac{AI}, as they contain not only coverage information but also sample coordinates, unlike \ac{MR} measurements, which lack georeferencing. In this case, the \ac{eNWDAF} would use \ac{MDT} data to train an \ac{ANN}, enabling it to infer the coordinates of \acp{MR} samples sent to the primary element by the \emph{AI-Agent} within the \acp{RBS}. This approach creates an optimal data provisioning scenario for \ac{ARCADE}, ensuring a high volume of information even if user adoption of \ac{MDT} functionality is low. Alternatively, in the absence of \ac{MDT} data for training the \ac{ANN}, information from alternative sources could be employed, such as proprietary solutions implemented by technology vendors or crowdsourced data made available by various app providers that collect such information.

\begin{figure}  
   \centering  
   \includegraphics[width=0.7\textwidth]{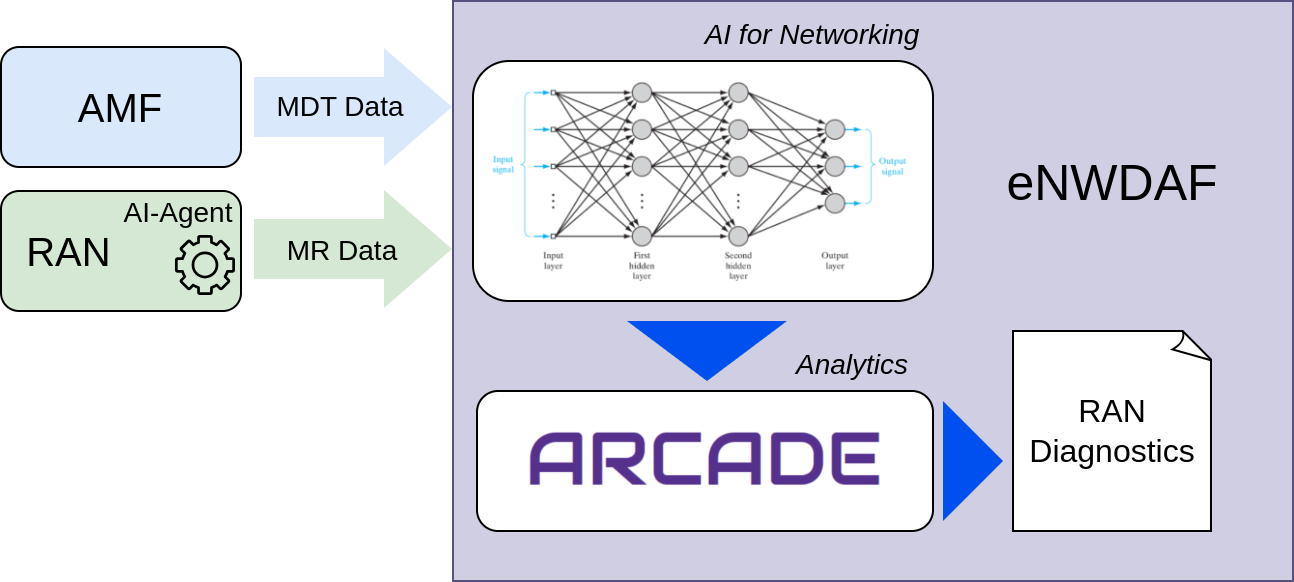}  
   \caption{Implementation of ARCADE within the hybrid architecture and eNWDAF concepts. Figure created by the authors.}  
   \label{fig:enwdaf_arcade}  
\end{figure}

This process clearly positions \ac{eNWDAF} within the role of \emph{Networking for AI} and Figure \ref{fig:enwdaf_arcade} illustrates how \ac{ARCADE} can be implemented within this hybrid approach to 6G network elements. It highlights its contribution to information acquisition and \ac{AI} application in the context of mobile network self-optimization, leveraging the concepts presented in this section.  

\section{Concluding Remarks}  
\label{sec:conclusoes}  

In this work, we introduced the concept of \ac{ARCADE}. We demonstrated how a cellular coverage diagnostic system can benefit from a hybrid approach to positioning \ac{AI} elements implemented centrally yet extending across all network segments, including the \ac{RAN}. Such an approach can be highly valuable in 6G systems, where \ac{AI} will be pervasive throughout the network, aiming to enhance automation, resource efficiency, and overall system and infrastructure management.

There are several ways to develop this proposal further. A practical implementation of a proof of concept for the \ac{eNWDAF}, its interfaces, and the \emph{AI-Agent} should be carried out as a first step. This should be followed by a validation phase using either simulated network data or a real-world dataset collected from a mobile operator. Additionally, the implementation of \ac{ARCADE} must be completed and integrated into the proof of concept, thereby establishing a comprehensive environment for testing and validating the proposed approach.

\section*{Acknowledgement}
The authors thank the support of 
FAPEMIG (Grant APQ00923-24) 
and FCT – Fundação para a Ciência e Tecnologia within the R\&D Unit Project Scope UID/00319/Centro ALGORITMI (ALGORITMI/UM) for supporting this work. The authors also thank Algar Telecom for their support in our research.

\bibliographystyle{sbc}
\bibliography{references}

\end{document}